\documentclass[prl, twocolumn, superscriptaddress, numerical]{revtex4-2}
\usepackage{palatino}
\usepackage[latin1]{inputenc}
\usepackage{epsf}
\usepackage{amsmath,amssymb}
\usepackage{latexsym}
\usepackage{calc}
\usepackage{color}
\usepackage{shadow}
\usepackage{epsfig}
\usepackage[usenames,dvipsnames]{xcolor}
\usepackage{subfig}

\begin{document}

\title{Probing particle-particle correlation in harmonic traps with twisted light}

\author{Johanna I. Fuks}
\affiliation{Departamento de F\'isica and IFIBA, FCEN, Universidad de Buenos Aires, Ciudad Universitaria Pabell\'on I, 1428 Ciudad de Buenos Aires, Argentina}
\author{Guillermo F. Quinteiro}
\affiliation{Instituto de Modelado e Innovaci\'on Tecnol\'ogica, and Departamento de F\'isica, FaCENA,
Universidad Nacional del Nordeste, 3400 Corrientes, Argentina}
\author{Heiko Appel}
\affiliation{Max Planck Institute for the Structure and Dynamics of Matter,
Luruper Chaussee 149, D-22761 Hamburg, Germany}
\author{Pablo I. Tamborenea}
\affiliation{Departamento de F\'isica and IFIBA, FCEN, Universidad de Buenos Aires, Ciudad Universitaria Pabell\'on I, 1428 Ciudad de Buenos Aires, Argentina}

\date{\today}

\begin{abstract}
We explore the potential of twisted light as a tool to unveil many-body effects in parabolically confined systems.
%via its non-homogenous spatial structure.
 %We explore the potential of twisted light, via its non-homogenous spatial structure, 
%as a tool to unveil many-body effects in parabolically confined systems. 
%
According to the Generalized Kohn Theorem, the dipole response of such a multi-particle system to 
a spatially homogeneous probe is indistinguishable from the response of a system of 
non-interacting particles.
%
%We find that the latter is also true for the quadrupole response.
Twisted light however can excite internal degrees 
of freedom, resulting in the appearance of new peaks in the even multipole spectrum 
which are not present when the probe is a plane wave. 
We also demonstrate the ability of the proposed twisted light probe 
to capture the transition of interacting fermions into a strongly 
correlated regime in a one-dimensional harmonic trap. 
We report that by suitable choice of the probe's parameters, the transition into 
a strongly correlated phase manifests itself as an approach and 
ultimate superposition of peaks in the second order quadrupole response. 
These features, observed in exact calculations for two electrons, 
are reproduced in adiabatic Time Dependent Density Functional Theory simulations.
%quadrupole spectrum.
\end{abstract}

\maketitle

Twisted light (TL), also known as optical vortices, designates a family of highly 
non-homogeneous optical beams which have single or multiple phase singularities and 
carry orbital angular momentum (OAM), among other interesting features \cite{Allen03,Andrews11}. 
%
%Already, 
Promising applications of TL have been identified in areas such as
telecommunications \cite{TLTelecomm17, TLTelecomm16}, 
quantum computing \cite{EFKZ18}, 
nanotechnology \cite{Cai12, Schulze17, Chen18}, 
and enhanced resolution imaging \cite{Ritsch17,TLPK17},
to name a few \cite{Mann18,Shen19}. 
From a fundamental point of view, researchers seek to understand the generation, 
detection, and interaction of TL with matter.
The latter is strongly affected by the structure of the light field, and the peculiar
features of TL have been shown to produce novel optical effects.
For example, rare transitions in atoms resulting from new selection rules \cite{Schmiegelow16, Schmiegelow18}, 
distinct time scales and lifetimes \cite{NK15} and degree of spin polarization for OAM exchange in GaAs \cite{SMSA19}, 
as well as coherent photo-exciton dynamics in GaN \cite{STM13} have been observed 
under TL excitation.
Based on recent progress a variety of new phenomena undetectable by plane waves, 
are expected to be revealed with TL.
Until now there has been an emphasis on studies of the interaction of TL with single 
particles, to the detriment of research contemplating the role of interactions \cite{QuinteiroCoulomb,Babiker18,HRK18}.
Here we investigate the response of multiparticle parabolically confined systems, notorious for their many-body effects going undetected when the probe is spatially homogeneous \cite{Brey89} .
We show theoretically key ways in which the many-body physics can be unveiled by TL.
%
%Harmonic traps can describe, among other systems, 
%interacting particles in parabolic quantum wells, 

Electrons in quantum wires and dots, 
%positively charged metal spheres, 
%Hooke's atoms 
%or spherical nuclear models
%by suitable choice of the matrix ${\bf K}$ 
ions in Paul traps and cold atoms in optical lattices can all be described by the same separable, parity conserving Hamiltonian \cite{Yip91}. For atoms the interaction is modeled as a contact potential \cite{BERW98}, whereas for electrons and ions the interaction is via a Coulomb potential 
\cite{Brey89,Deng07}.
The correlation regime of such harmonic traps can be tuned by varying the natural frequency $\omega_0$ of the confining parabolic potential and the number $N$ of trapped particles.

In virtue of the high tunability and the availability of analytic (weak and strong interaction limits)
\cite{Taut93, Taut03, BERW98} or very accurate  
one-dimensional numerical solutions \cite{Gharashi13, Grining15},
these systems provide a good testbed for many-body physics. 
Even more, harmonic traps can be realized experimentally and can be used as quantum simulators to study emergent many-body phenomena 
such as superconductivity, superfluidity, quantum phase transitions and topological order \cite{Georgescu14, Tarruell18}.

The Hamiltonian describing a harmonic trap containing $N$ particles is separable into a center of mass (CM) and a part that depends on the internal degrees of freedom,
$H_0({\bf r}_1..{\bf r}_N,{\bf p}_1..{\bf p}_N)=H_0^{\rm CM}({\bf R},{\bf P}_R)+ H_0^q({\bf q}_1...{\bf q}_{N-1},{\bf p}_{q_1}..{\bf p}_{q_{N-1}})$. 
The CM system is equivalent to a single particle with  mass $M=Nm$ and charge $Ne$
 \begin{equation}
 H_0^{\rm CM} = \frac{{\bf P}_R^2}{2M} + \frac{1}{2} M \omega_0^2{\bf R}^2,
 \label{eq:H0_CM}
 \end{equation}
 with  ${\bf R}=\frac{1}{N}\sum_i^N {\bf r}_i $ and ${\bf P}_R=\sum_i^N {\bf p}_i$.
For $N=2$
%(Hooke's atom if the particles are electrons) 
and definitions 
\mbox{${\bf q}=({\bf r}_1-{\bf r}_2)/\sqrt{2}$ and ${\bf{p}}_q=({\bf p}_1-{\bf p}_2)/\sqrt{2}$} for the Jacobi coordinates the internal Hamiltonian reads
\begin{equation}
     H^{q}_{0}= \frac{{\bf p}_q^2}{2m} + \frac{1}{2} m \omega_0^2{\bf q}^2 +  V_{\rm int}({\bf q})
     \label{eq:H0_rel}
\end{equation}
with $V_{\rm int}({\bf q})$ being the interaction between particles. 
Other choices of normalization are valid as long as the conmutation relations 
    $[{\bf R},{\bf P}_R]=i\hbar$ and $[{\bf q}_j,{\bf p}_{q_j}]=i\hbar$ hold.
%Psi({\bf r}_1,..,{\bf r}_N},{\bf p}_1,..,{\bf p}_N})= \phi({\bf R},{\bf P}_{R}) \varphi(\bf{q}_1,..,\bf{q}_{N-1},\bf{p}_{q_1},.., \bf{p}_{q_{N-1}}).

 A homogeneous light field (plane waves with long wave length) couples to the dipole moment of the system (Dipole Approximation), which is a purely 
 CM variable, 
  and can therefore only induce  transitions between states of the CM Hamiltonian. This is the principle supporting the predictions of the Generalized Kohn Theorem 
  (GKT) \cite{Brey89, Yip91}
  and also of the Harmonic Potential Theorem (HPT) \cite{D94}. 
  GKT proves that by means of a spatially homogeneous probe, no many-body effects can be observed
  in the dipole response.
Whereas for non-harmonic traps signatures of particle-particle correlation can be observed in the dipole spectrum 
\cite{Escartin12}, in separable systems the probe 
needs to be non-homogeneous to be able to excite internal transitions
\cite{M98,WCM95}. 
We propose to use the spatial inhomogeneity associated with the topological charge $l$ of the TL field to study 
the internal energy spectrum of the trap.

The quantum number $l$
characterizes the OAM the TL beam transports.
A phase singularity 
associated with $l$ gives rise to a spatially non-homogeneous and hollow intensity distribution with vanishing electromagnetic fields 
at the beam center. 
In the Coulomb gauge, ${\nabla \cdot}{\bf A}^{Cou}({\bf r},t)=0$, and the scalar potential can be chosen to vanish, $\Phi^{Cou}=0$.
%and within the paraxial approximation (PA), 
We can then write the vector potential of a monochromatic TL field in cylindrical coordinates as \cite{QuinteiroBessel19}
\begin{equation}
  %\begin{split}
  {\bf A}({\bf r}, t)
  %=  F_{q_rl}(r)
  =F_{q_rl}(r) e^{i \theta }e^{il\varphi} {\bf \epsilon_\sigma}
  %({\bf \epsilon_x} -i \sigma {\bf \epsilon_y})/\sqrt{2}
  %+ c.c)
     - i\sigma \frac{q_r}{q_z} \frac{F_{q_rl+\sigma}(r)}{\sqrt{2}} e^{i \theta} e^{i(l+\sigma)\varphi} {\bf \epsilon_z}   + c.c.
%F_{q_rl+\sigma
     %\end{split}
  \label{eq:A_TL}
\end{equation}
with $\theta=q_z z-\omega t$ and frequency given by $\omega^2=c^2(q_r^2+q_z^2)$, where $1/q_r$ is a measure of the beam waist.
The radial profile is described by a Bessel function $F_{q_rl}(r)=A_0 J_l(q_r r)$, 
the polarization vector is given by
%/(2^l l!)$.
  ${\bf \epsilon}_{\sigma}=e^{i\sigma \varphi}({\bf r} + i \sigma \varphi)/\sqrt{2}=({\bf x} + i\sigma {\bf y})/\sqrt{2}$.
To describe the TL-matter interaction 
we choose the TL gauge introduced in ref.~\cite{QRK15}. 
In the TL gauge the interaction with a small, planar and localized structure placed close to the phase singularity can be written 
in a gauge invariant form.
For a parallel TL beam, sign($\sigma$)=sign($l$), with circular polarization $\sigma=1$,
$H_{I}^{TL} = \frac{-e}{l+1} {\bf r}_{\perp}\cdot \frac{\partial}{\partial_t} {\bf A}$. 
\begin{equation}
%\begin{split}
 H_{I}^{TL} 
 %= \frac{q}{l+1} {\bf r}_{\perp}\cdot \frac{\partial}{\partial_t} {\bf A}^{PA}
% \\&=\frac{q}{l+1}{\bf r}_{\perp}\cdot  \frac{\partial}{\partial_t}
 %\big(A_0 e^{il\varphi}e^{i(q_z z-\omega t)} ({\bf x} +i {\bf y})(q_r r)^l  + c.c.\big)
 =  -\frac{i e \omega F_{q_r l}(r)}{(l+1)}
 %(q_r r)^l
 [ -(x+iy) e^{i \theta}
 %{i(l\varphi -\omega t)}
 +(x-iy) e^{-i \theta}]
 %{-i(l \varphi -\omega t)} ]
 \label{eq:Hint_TLgauge}
%\end{split}
 \end{equation}
 where $e$ is the electron charge,
${\bf r}_{\perp}\cdot({\bf x} +i {\bf y})=x+iy$ and we have assumed $q_r\gg q_z$,
  % where PA stands for paraxial approximation,   
  which in the Coulomb gauge corresponds to neglecting the second term in Eq.~(\ref{eq:A_TL}).
Making use of the equivalence { $r^le^{i l \varphi}=(x+iy)^l$ } and approximating $J_l(q_r r)\approx (q_r r)^l/(2^l l!)$ near ${\bf r}=0$ we 
arrive at an expression
in which the interaction with an $N$-particle compact object placed at $z=0$ is described by a scalar field which for $l=1,2$ reads
\begin{equation}
%H_{I}^{l=1}=\frac{-e E_0q_r}{4} \sum_i^N {(-x_i^2 + y_i^2)sin(\omega t) + 2x_i y_i \cos(\omega t)},
H_{I}^{l=1}=\frac{e E_0q_r}{2 \sqrt{2}} \sum_{i=1}^N {(x_i^2 - y_i^2)\sin(\omega t) - 2x_i y_i \cos(\omega t)},
 \label{eq:scalarTLl1}
 \end{equation}
%and for $l=2$,  
\begin{equation}
%H_{I}^{l=2}=\frac{e E_0 q_r^2}{24} \sum_i^N(x_i^3-3x_iy_i^2) sin (\omega t) + (y_i^3-3y_ix_i^2) \cos(\omega t ),
H_{I}^{l=2}=\frac{e E_0 q_r^2}{12 \sqrt{2}} \sum_{i=1}^N(x_i^3-3x_iy_i^2) \sin (\omega t) + (y_i^3-3y_ix_i^2) \cos(\omega t ),
 \label{eq:scalarTLl2}
\end{equation}
where $E_0=\omega A_0$. When $l=0$ the field couples to the dipole (dipole approximation).
For $N=2$ and one spatial dimension we can rewrite Eqs.~(\ref{eq:scalarTLl1})-(\ref{eq:scalarTLl2}) in terms of CM and Jacobi coordinates as,
\begin{equation}
%\begin{split}
 %H_{I}^{l=1}= -eE_0 q_r (2X^2 +q^2)\sin(\omega t)
 H_{I}^{l=1}= \frac{eE_0 q_r}{2\sqrt{2}} (2X^2 +q^2)\sin(\omega t),
 %\\ &  + 2(2YX +q_yq_x) cos(\omega t))
 %\end{split}
 \label{eq:scalarTLl1_cm_q}
 \end{equation}
 \begin{equation}
% \begin{split}
%H_{I}^{l=2}=\frac{4e E_0 q_r^2}{3} (2X^3+Xq^2) \sin (\omega t) 
H_{I}^{l=2}=\frac{e E_0 q_r^2}{12\sqrt{2}} (2X^3+Xq^2) \sin (\omega t).
 %\\ & + ((4Y^3+2Yq_y^2)-6(2YX^2 + Yq_x^2)) cos(\omega t ).
 %\end{split}
 \label{eq:scalarTLl2_cm_q}
\end{equation}
  Notice that for $l=1$ the interaction with TL is separable into  CM and internal parts;
%{\it check that this is indeed true for $N=3$, what about $N>3$?} 
for $l\geq2$ however, separability breaks and the system's response can no longer be described as the sum of the responses of CM and internal systems. 
Hamiltonians Eqs.~(\ref{eq:H0_CM})-(\ref{eq:H0_rel}) conserve parity,
therefore  all transitions between even and odd states are forbidden within first order perturbation theory. 
For a sufficiently small applied field, the response can be expanded in a power series of the strength $E_0$.
The linear ($\propto E_0$) and second order ($\propto E_0^2$) responses 
of an observable ${O}$ are proportional to the matrix elements
\mbox{$\langle \hat{O}^{(1)}\rangle \propto  \langle \Psi_0|\hat{O}|\Psi_k\rangle \langle\Psi_k|\hat{H}_I|\Psi_0 \rangle$} and 
\mbox{$\langle \hat{O}^{(2)}\rangle \propto \langle \Psi_0|\hat{O}|\Psi_k\rangle \langle\Psi_k|\hat{H}_I|\Psi_m \rangle\langle\Psi_m|\hat{H}_I|\Psi_0 \rangle$} 
respectively \cite{Fetter12, gvbook2005, IYB01, Gray76}.
As a consequence, only if the interaction and the observable operators have the same parity will the response be first order in the perturbation, 
otherwise the first non-vanishing response will be second order. 
 In Table \ref{tab:response} we list the dipole ($D=\langle -e\sum_i^N\hat{x}_i\rangle$) and quadrupole ($Q=\langle -e\sum_i^N\hat{x}_i^2\rangle$) responses 
 of a one-dimensional harmonic trap perturbed with a TL probe for three different spatial dependences of the probe ($l=0,1,2$).
 %for a small trap placed close to the beam center. 
 The frequencies of the observable transitions are shown inside square brackets. 
 %The first row, first column 
 $D_{\rm CM}^{(1)}[(2 n+1)\omega_0]$
 %in Table \ref{tab:response} 
corresponds to GKT:
 for an $l=0$ homogeneous probe the dipole response is equal to the CM response and first order in the field,
 with poles at the odd CM transitions $\omega^{\rm CM}_{0,(2n+1)}=(2n +1)\omega_0$.  
 We focus on the quadrupole response  because we are interested in correlation effects and the dipole response can only observe the CM spectrum (see first column Table \ref{tab:response}). 
 \begin{table}
   \begin{center}
   \caption{Selection rules one-dimensional two-particle harmonic-trap response to TL probe. Poles in brackets.}
  \begin{tabular}{ |c | c | c |  }
 \hline
   observable: &$\hat{D}=-e\hat{X}$ &   $\hat{Q}=-e(2 \hat{X^2}+\hat{q^2})$ 
  \\
  \hline
   $\hat{H}_{I}^{l=0}$ &   $D^{(1)}_{\rm CM}$ [$(2n+1)\omega_0$]  &   $Q^{(2)}_{\rm CM}$ [$2n\omega_0$]     
 \\
  \hline
  $\hat{H}_{I}^{l=1}$  &  $D^{(2)}_{\rm CM}$ [$(2n+1)\omega_0$]  &   $Q^{(1)}_{\rm CM}$ [$2n\omega_0$] 
   ;  $Q^{(1)}_{q}$ [$\omega^q_{0 2m}$]  
 \\
   \hline
    $\hat{H}_{I}^{l=2}$ &  $D^{(2)}_{\rm CM}$ [$(2n+1)\omega_0$]  &  $Q^{(2)}_{\rm CM}$ [$2n\omega_0$] 
   ;  $Q^{(2)}_{q}$ [$\omega^q_{2m 2(m+1)}$]  
 \\
 \hline 
 \end{tabular}
 \label{tab:response}
\end{center}
 \end{table}

In order to evaluate whether a TL probe is able to capture relevant many-body physics we investigate the transition into a strongly correlated regime, namely the limit $\omega_0\rightarrow 0$.
We model the interaction between particles with a soft-Coulomb potential \cite{L16,G17}
%{HFTAGR11} 
%We model the Coulomb interaction 
\begin{equation}
     %V ({x}_i-{x}_j)=\frac{e^2}{\sqrt{(x_i-x_j)^2+a^2}}.
     V_{int}(q)=\frac{e^2}{\sqrt{(\sqrt{2}q)^2+a^2}}.
    \label{eq:SoftCou}
\end{equation}
%$\alpha$ is the strength of the particle-particle interaction and $a$ is the soft-Coulomb parameter. 
We use atomic units (a.u.) for all the calculations, $\hbar=e=m_e=a_0=1$, and take $a=1$. For a semiconductor the effective units of length, energy and time transform as $a_0^*=(\epsilon/m^*)a_0$, ${Ha}^*=(m^*/\epsilon^2)Ha$ and $u_t^*=(\epsilon^2/m^*)u_t$, where $\epsilon$ and $m^*$ are the relative permitivity and effective mass of the material. 
 \begin{figure}
\includegraphics[width=0.49\textwidth]
 %{../after_Heikos_fix/Dirac/plots/
 {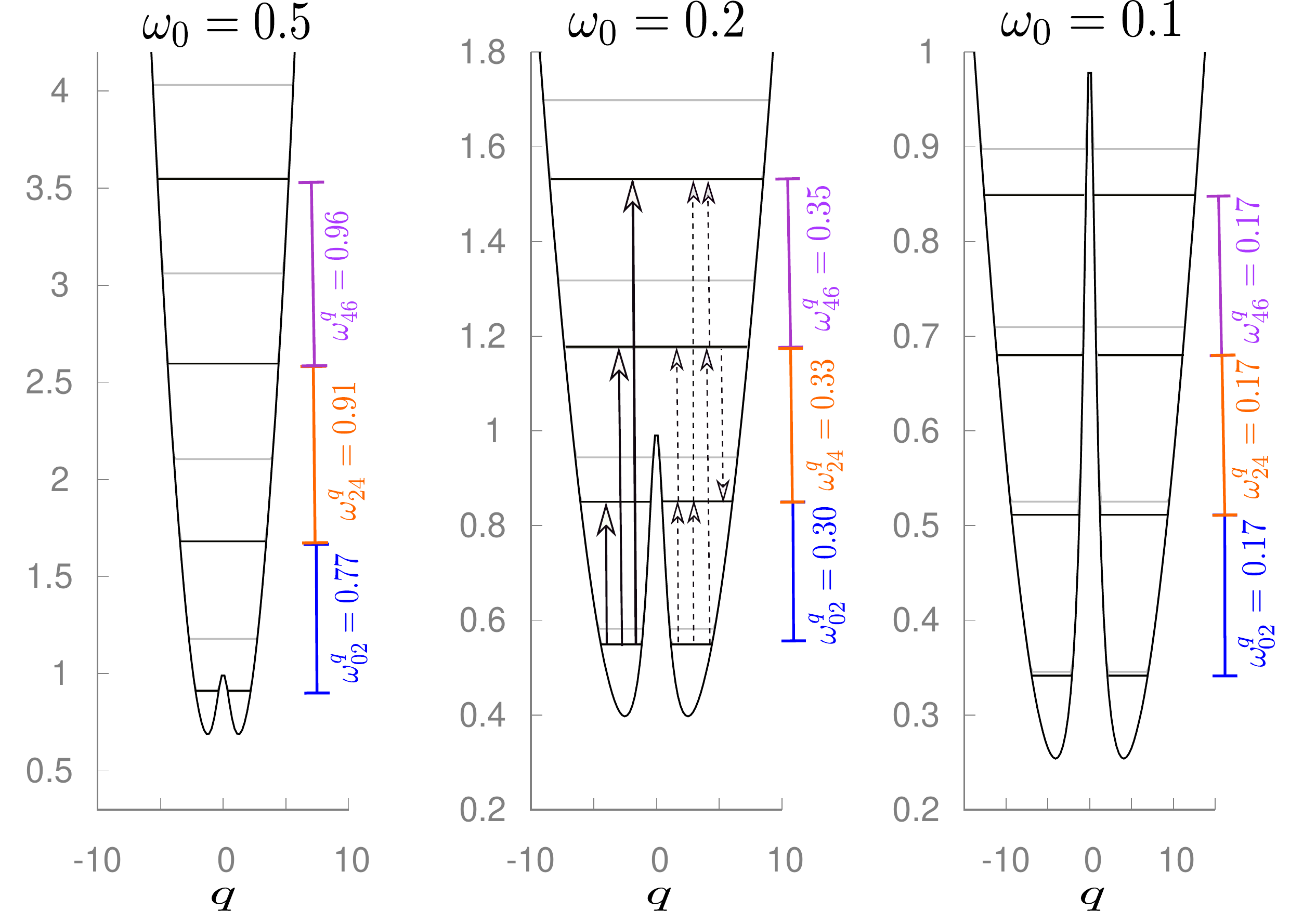}
                %{../after_Heikos_fix/Dirac/plots/V_vs_q_ev_q_diff_w0_8evs_greyBorders.pdf}
\caption{Internal potential $V(q)$, eigenvalues
%$1/2 m\omega_0^2 q^2 + V(q)$ 
%for $\omega_0=0.5$ (left), $\omega_0=0.2$ (middle) and $\omega_0=0.1$ (right),
$\epsilon^q_n$ and energy spacings $\omega^q_{mn}=\epsilon^q_m-\epsilon^q_n$ in the low energy region (atomic units).
The solid(dashed) arrows represent the first order $Q_q^{(1)}$ (second order  $Q_q^{(2)}$ ) quadrupole response to a quadrupole probe.
}
                \label{fig:fig2}
\end{figure}

\begin{widetext}
\begin{figure*}
\includegraphics[width=1.0\textwidth]
                %{../after_Heikos_fix/Dirac/plots/
                {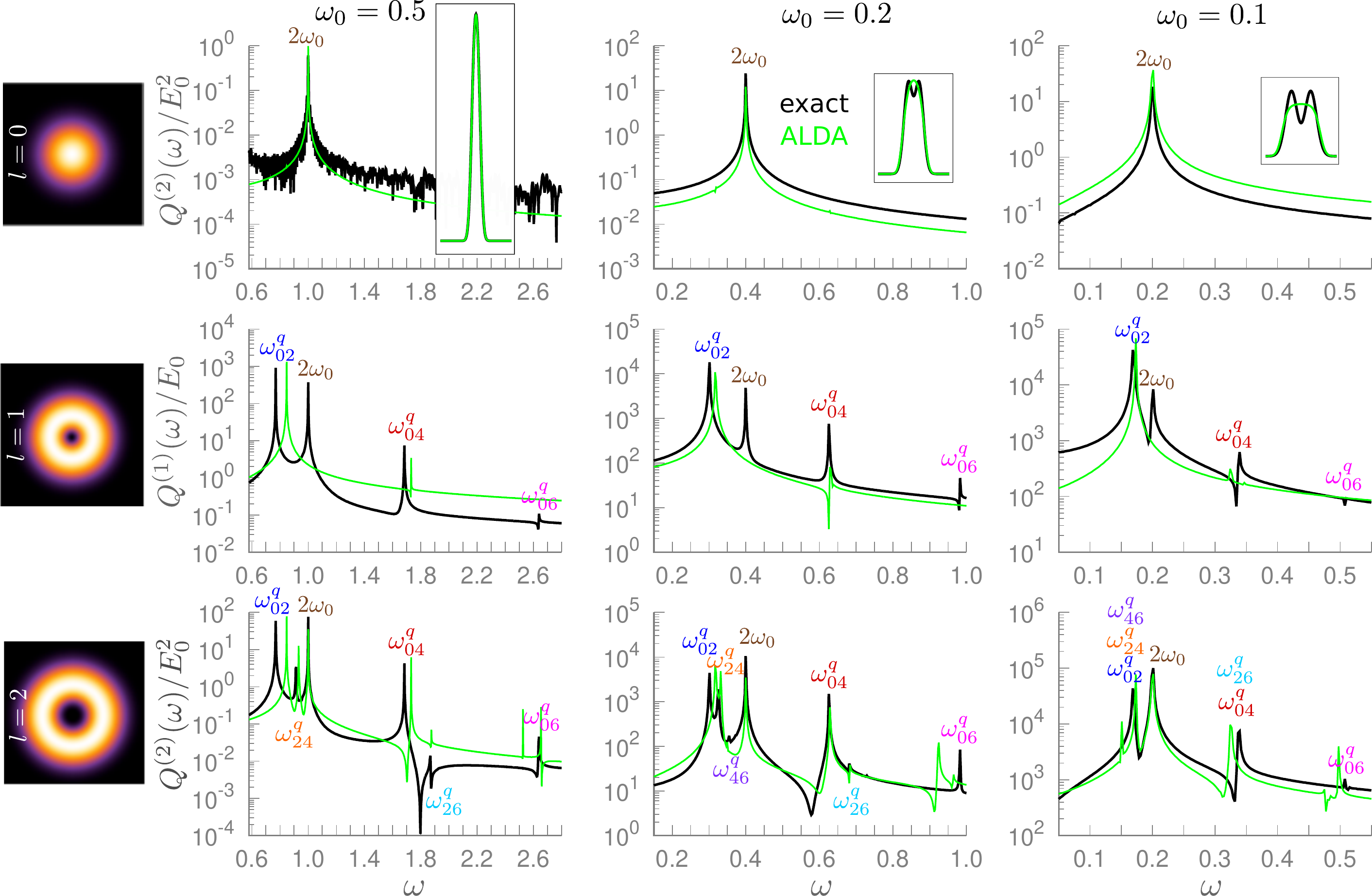}
\caption{Exact (black) and ALDA (green) quadruple spectrum %$\langle \sum_i x_i^2\rangle(\omega)/E_0^{(n)}$ 
of a two-electron one-dimensional harmonic trap excited with  
an  homogeneous $l=0$ probe (upper panel), a $l=1$ TL probe (middle panel) and a 
$l=2$ TL probe (lower panel). The ground state densities are shown in the inset. Numerical values for
$\omega^q_{mn}$ in Fig.~\ref{fig:fig2} (atomic units).
}
                \label{fig:fig3}
\end{figure*}
\end{widetext}

To identify the features that characterize the strongly correlated regime we compute the 
energy spectrum of a one-dimensional two-particle harmonic trap 
for various confinement strengths $\omega_0$. 
In Fig.~\ref{fig:fig2} we represent the internal potential \mbox{$V(q)=1/2 m\omega_0^2 q^2 + V_{\rm int}(q)$}, first 8 eigenergies $\epsilon^q_k$ and first 3 energy spacings $\omega^q_{mn}=\epsilon^q_m-\epsilon^q_n$ for 3
different confinement strengths $\omega_0$. We identify several features that characterize the limit $\omega_0\rightarrow 0$ :
i) The particles localize maximally far from each other. This 
is evident from the shape of $V(q)$, which transitions from a single to a double well. This behaviour is also reflected in the shape of the total density (see inset Fig.~\ref{fig:fig3}) and is usually referred to as low density limit or Wigner crystal in the literature \cite{PKSRRMI13,VDS20,MGG18,Rasanen03} and .
ii) Increasing deviation of the ground state wavefunction from a Single Slater Determinant (strong correlation),
  resulting in an increment of the 
 van Neumann entropy \cite{Z95}: $s^{\omega_0=0.5}=0.07$; $s^{\omega_0=0.2}=0.25$; $s^{\omega_0=0.1}=$ 0.35. 
iii) The symmetric and antisymmetric wavefunctions 
 become energetically degenerate 
 \cite{Guan2009} as in the case of distinguishable fermions \cite{CJM20,SGDL13}, which can be seen in the approaching of even and odd internal levels (black and grey in Fig.~\ref{fig:fig2})
 \footnote{For separable systems it is the parity of the internal wavefunction that determines the particle exchange symmetry and therefore the symmetry of the spin wavefunction \cite{LandauQM65}.}.
 iv) 
 The internal energy levels become equidistant
 %$\omega^q_{n,2n}\rightarrow \sqrt{3}\omega_0$ 
 as in the case of a single-particle harmonic oscillator; with an effective natural frequency consistent with $\omega^q_{n,n+1}\rightarrow \frac{\sqrt{3}\omega_0}{2}$ as $\omega_0\rightarrow 0$. This result is consistent with a Taylor expansion of the Coulomb potential 
 %In the limit of strong correlation the Coulomb potential can be expanded 
 around the classical equilibrium positions of the particles \cite{Taut93,JH96}.
 We show that we can identify, by means of a TL probe, this characteristic equispacing in the quadrupole response.

The arrows in Fig.~\ref{fig:fig2} represent the quadrupole-allowed internal transitions in the low energy region. Solid arrows connect the ground state with even states and represent the first order quadrupole response 
$Q^{(1)}_q$, these transitions can be excited with a TL probe of $l=1$. Dashed arrows correspond to $Q^{(2)}_q$ which depends on two frequencies (two-photon processes). These transitions connect even internal states and can be excited with a $l=2$ TL probe (see Table \ref{tab:response}). 
In Fig.~\ref{fig:fig3} we present the exact (black) and Adiabatic Local Density Approximation (ALDA) (green) quadrupole spectrum for TL probes of $l=0,1,2$. What we plot is the absolute value of the Fourier Transform (FT) of the quadrupole, \mbox{$Q(\omega)=-e\langle\sum_i^{N=2}\hat{x}_i^2\rangle(\omega)=FT[ -e\int n(x,t) \hat{x}^2 dx]$},  
%{\it absorption exp?ref?}.
computed from real-time evolution of the electronic density $n(x,t)$. The TL probe is modeled as a weak impulse field, $H_{I}^{l}=E_0\sum_i^N x_i^{(l+1)}\delta(t)$. Because TL-matter interaction in the case under study is modeled as a scalar potential, the response can be written in terms of the density-density response and can be simulated with TDDFT \cite{TDDFTbook12}. 
The imaginary parts of $Q^{(1)}(\omega)$ and $Q^{(2)}(\omega)$ correspond to the quadrupole polarizability and  first quadrupole hiperpolarizability respectively \cite{Buckingham79, Higgins20}, which can be measured in an absorption experiment \cite{Tojo04,Ray20}. 
For the homogeneous probe $l=0$ shown in upper panel the quadrupole response is of second order and purely CM as predicted in Table \ref{tab:response}, with one unique visible peak  
at  $\omega^{\rm CM}_{2n,2(n+1)}=2\omega_0$.
ALDA reproduces the position of the peak accurately.
For the $l=1$ TL probe shown in middle panel the quadrupole response is of first order. Four 
peaks are visible in the low energy region, corresponding to the transitions $\omega^q_{02}$, $\omega^{\rm CM}_{02}=2\omega_0$, $\omega^q_{04}$ and  $\omega^q_{06}$ indicated with solid arrows in Fig.~\ref{fig:fig2}. The numerical values of the internal transition frequencies $\omega^q_{nm}$ are also shown in Fig.~\ref{fig:fig2}.  As correlation grows the internal transitions become equidistant and independent of the interaction, $\omega^q_{2n,2(n+1)}\rightarrow \sqrt{3} \omega_0$.
%as predicted theoretically for the limit $\omega_0\rightarrow 0$ \cite{Taut93,Taut03}.
%as predicted theoretically from a Taylor expansion of the Coulomb term around the classical equilibrium position of th \cite{Taut93, Taut03}.
ALDA shows one unique peak instead of two for $l=1$; this shortcoming is because double excitations cannot be modeled within linear response Adiabatic TDDFT \cite{TDDFTbook12}. 
In ref.~\cite{MZCB04} an ad hoc correction
that splits the peak into two was proposed and accurate energies for the transitions $\omega^q_{02}$ and $\omega^{\rm CM}_{02}$ were obtained for the same one-dimensional Hooke's atom 
%two-electron harmonic trap 
studied here.
%and the corrected frequencies are shown with arrows {\it compute correction!!!!!}.
For the $l=2$ TL probe shown in the lower panel the quadrupole response is of second order and shows peaks at 
%, with contributions from both CM and internal transitions at  
%the peaks  presents peaks at the same frequencies as $l=1$ TL probe 
$\omega^q_{02}$, $\omega^{\rm CM}_{02}=2\omega_0$, $\omega^q_{04}$ and $\omega^q_{06}$ as for $l=1$, plus additional peaks at $\omega^q_{24}$, $\omega^q_{46}$ and $\omega^q_{26}$ corresponding to the transitions indicated with dashed arrows in Fig.~\ref{fig:fig3}. 
%Transitions $\omega^q_{2n,2(n+1)}$ with $n>2$ also lye in this region but since their oscillator strengths are small higher resolution is required to 
 As we move into the strongly correlated regime the transition frequencies $\omega^q_{02}$, $\omega^q_{24}$ and $\omega^q_{46}$ get closer and so do $\omega^q_{04}$ and $\omega^q_{26}$. For  $\omega_0=0.1$ a.u.. they overlap at the theoretical value $\omega^q_{2n,2(n+1)}\rightarrow \sqrt{3} \omega_0$ and $\omega^q_{2n,2(n+2)}\rightarrow 2 \sqrt{3} \omega_0$ respectively, as predicted theoretically for the limit $\omega_0\rightarrow 0$ \cite{Taut93, Taut03}.
ALDA reproduces pretty well the position of the peaks in the $l=2$ spectra for all $\omega_0$, capturing the approaching and overlapping of the peaks as the confinement $\omega_0$ decreases and correlation dominates. Despite reproducing poorly the ground state density in this limit \cite{MGG18} (see inset Fig.~\ref{fig:fig3}) ALDA's spectrum seems to get actually more accurate for small $\omega_0$. Maybe the equidistant spacing of the internal energy levels is easier for the TDDFT approximation to capture because it renders the response effectively single-particle. Nevertheless, for $\omega_0=0.1$ we observe an additional peak in the low energy region of the ALDA spectrum. This spurious peak coincides with the energy difference between the second and fourth ALDA peaks. The Exact Exchange (EXX) approximation does not improve the spectra nor the ground state density.
Both the exact and TDDFT dynamics were computed in the octopus code \cite{octopus2012, HFTAGR11} using a  
$100$ a.u. simulation box with $0.1$ a.u. spacing, 
%was used and for the three electron problem $20$ a.u. and $0.2$ a.u. respectively. 
a $0.005$ a.u. time step and a total propagation time $T=2500$ a.u., leading to a peak resolution of $2\pi/T\approx0.0025$ a.u..

We have shown analytically that, unlike plane waves, a TL field is able to excite internal transitions in separable systems of interacting particles. 
We identify several features that characterize the transition into a strongly correlated regime in a many-particle harmonic trap. Out of them the degeneracy between symmetric and antisymmetric wavefunctions has been experimentally observed in nanowires and in a one-dimensional atom trap \cite{PKSRRMI13,Zuern12}. We show that another characteristic feature, namely the equidistance between internal energy levels, can be revealed in the quadrupole response after excitation with a TL probe.
%We show the transition into a strongly correlated phase manifests in the quadrupole spectrum and 
We present numerically exact calculations on a two particle quantum wire and show that
the results can be reproduced with an ab initio method such as adiabatic TDDFT.
The validity of our findings is expected to hold for an arbitrary number of particles, as supported by ALDA simulations for $N=3$.
We expect that in the case of a two-dimensional harmonic trap,
%for which the interaction with the TL field is aldescribed by Eqs.\ref{eq:scalarTLl1}-\ref{eq:scalarTLl2} hold, 
the quadrupole response contains information on the moment of inercia; this may prove useful to study angular momentum exchange \cite{Babiker02} or superfluidity \cite{ZS01}. 
We stress that the scheme we are proposing is useful to study the transition into a strongly correlated regime in quantum wires or wells but also in one or two-dimensional Paul traps containing several ions \cite{Oral19}. Given the experimental feasibility and high tunability of both TL beams and harmonic traps it could also serve as a quantum simulator for intermediate correlation regimes for which no analytic solutions exist.
Alternatively it could also be used to prepare the system in an 'optically active' state of choice or even to characterize the TL field. Whether a TL probe can provide additional information about the degree of correlation in the more general case of non-harmonic systems for which GKT does not apply is the matter of future work.

\acknowledgments

JIF and PIT gratefully acknowledge the financial support from the Universidad de Buenos Aires, project UBACyT 2018-2020 No. 20020170100711BA. GFQ aknowledges ONRG through grant N62909-18-1-2090.

%

%\bibliographystyle{apsrev4-2}
%{bst}
%\bibliography{ref_exp}

\end{document}